\documentclass[narrowdisplay,onecolumn]{elsart3p}

\usepackage{graphicx}
\usepackage{amssymb}
\usepackage{lineno}
\usepackage{slashed}

\newcommand{\trace}{\mbox{tr}}  % defines the trace operator
\newcommand{\bra}[1]{\left< #1 \left|}  % defines the Bra
\newcommand{\ket}[1]{\right| #1 \right>}        % defines the Ket
\newcommand{\chiral}[1]{\mathring{#1}} % places a "chiral" circle on top of a symbol
\newcommand{\eqref}[1]{(\ref{#1})}
\newcommand{\Order}{\mathcal{O}}

\begin{document}
\hfill {\tiny HISKP-TH-09/32, FZJ-IKP-TH-2009-33}

\begin{frontmatter}
\title{New insights into the neutron electric dipole moment}
\author[1]{K. Ottnad},
\author[1]{B. Kubis},
\author[1,2]{U.-G. Mei\ss ner},
\author[2]{F.-K. Guo}
\address[1]{Helmholtz-Institut f{\"u}r Strahlen- und Kernphysik (Theorie) and
  Bethe Center for Theoretical Physics\\
  Universit{\"a}t Bonn, D-53115 Bonn, Germany}
\address[2]{Institut f{\"u}r Kernphysik, J\"ulich Center for
  Hadron Physics and Institute for Advanced Simulation\\
Forschungszentrum J{\"u}lich, D-52425 J{\"u}lich, Germany}

\begin{abstract}
We analyze the CP-violating electric dipole form factor of the nucleon
in the framework of covariant baryon chiral perturbation theory. 
We give a new upper bound on the vacuum angle, $|\theta_0| \lesssim 2.5 \cdot 10^{-10}$.
The quark mass dependence of the electric dipole moment is discussed and compared to
lattice QCD data. We also perform the matching between its representations in
the three- and two-flavor theories.
\end{abstract}
\end{frontmatter}

%%%%%%%%%%%%%%%%%%%%%%%%%%%%%%%%%%%%%%%%%%%%%%%%%%%%%%%%%%%%%%%%%%%%%%%%%%%%%%%
{\bf 1.} The neutron electric dipole moment (nEDM) is a sensitive probe of CP
violation in the Standard Model and beyond. The current experimental
limit $d_n \leq 2.9 \cdot 10^{-26}\, e \, {\rm cm}$~\cite{Baker:2006ts}
is still orders of magnitude larger than
the Standard Model prediction due to weak interactions. However, in QCD 
the breaking of the $U(1)_A$ anomaly allows for strong CP violation,
which is parameterized through the vacuum angle $\theta_0$. Therefore,
an upper bound on $d_n$ allows to constrain the magnitude of  $\theta_0$. 
New and on-going experiments with ultracold neutrons
strive to improve these bounds even further, see e.g.~\cite{LaGo} for 
a very recent review. On the
theoretical side, first full lattice QCD calculations of the neutron
electric dipole moment  are becoming available~\cite{Berruto:2005hg,Shintani:2008nt,Aoki:2008gv}.
These require a careful study of the quark mass dependence of the nEDM
to connect to the physical light quark masses. In addition, CP-violating
atomic effects can be sensitive to the nuclear Schiff moment, which 
receives a contribution from the radius of the nucleon electric dipole
form factor, see e.g.~\cite{Thomas:1994wi}.
It is thus of paramount interest to improve the existing
calculations of these fundamental quantities in the framework of 
chiral perturbation theory. In~\cite{Borasoy:2000pq}, the electric 
dipole moments of the neutron and the $\Lambda$ were calculated within the 
framework of $U(3)_L\times U(3)_R$ heavy-baryon chiral perturbation theory 
and an estimate for  $\theta_0$ was given (for earlier works utilizing 
chiral Lagrangians, see~\cite{Crewther:1979pi,Pich:1991fq}). 
In~\cite{Hockings:2005cn}, the
electric dipole form factor of the nucleon was analyzed to leading one-loop
accuracy in chiral $SU(2)$, thus in that calculation the form factor
originates entirely from the pion cloud. The strength of the form factor
was shown to be proportional to a non-derivative, time-reversal-violating
pion--nucleon coupling $\bar g_{\pi NN}$ that could only be estimated from dimensional analysis.
Furthermore, the leading contributions to the nEDM at finite volume and in
partially-quenched calculations were considered 
in~\cite{O'Connell:2005un}, and in~\cite{Chen:2007ug} the leading 
order extrapolation formula using a mixed action chiral Lagrangian is given.
In this Letter, we extend the results 
of~\cite{Borasoy:2000pq,Hockings:2005cn} to higher order based on a
covariant version of $U(3)_L\times U(3)_R$ baryon chiral perturbation theory.
This allows to make contact to the lattice QCD results 
from~\cite{Shintani:2008nt} and by matching, we can also get more insights
into the nucleon electric dipole form factor and the size of the 
coupling constant $\bar g_{\pi NN}$.

\medskip
\noindent
{\bf 2.} The electromagnetic form factors of the nucleon 
(in the presence of P and CP violation) are defined by means of the corresponding
three-point function
\begin{equation}
 \bra{p'} J_{em}^{\nu}\ket{p} 
= \bar{u}\left(p'\right)\Gamma^{\nu}\left(q^2\right) u\left
(p\right) \ ,
 \label{eq:EMCurrent}
\end{equation}
where $J_{em}^{\nu}$ denotes the  electromagnetic current and
\begin{equation}
 \Gamma^{\nu} = \gamma^{\nu}F_1\left(q^2\right) - \frac{i}{2m}\sigma^{\mu\nu}q_\mu F_2\left(q^2\right) 
+ i \left( \gamma^{\nu}q^2 - 2m q^{\nu}\right) \gamma_5 F_A\left(q^2\right) 
- \frac{1}{2m} \sigma^{\mu\nu}q_\mu\gamma_5 F_3\left(q^2\right)
\ ,
\label{eq:formfactors}
\end{equation}
with $q_\mu = (p'-p)_\mu$.
Here, $F_1$ and $F_{2}$ denote the P-, CP-conserving Dirac and Pauli 
form factors, and $m$ is the mass of the nucleon.
The last two form factors $F_{A}$ and $F_3$ 
stem from P- and CP-violating terms, respectively:  $F_{A}$
denotes the anapole form factor and $F_3$ the electric dipole form factor. 
In what follows, we will only consider the dipole form factor $F_3$.
The electric dipole moment of the neutron/proton is defined as the electric 
dipole form factor at $q^2=0$ 
\begin{equation}
d_{n,p} = \frac{F_{3,n,p}(0) }{2m}~.
\label{eq:EDM_def}
\end{equation}
Expanding the form factor in the squared momentum transfer allows one to
define an electric dipole radius,
\begin{equation}
 \left\langle r_{ed}^2 \right\rangle 
= 6 \frac{d F_3(q^2)}{dq^2}\bigg|_{q^2=0} \ .
\label{eq:EDMR_def}
\end{equation}
Note that similar to the case of the neutron electric form factor, we do
not include the normalization of the form factor at $q^2 = 0$ in this 
definition.

\medskip
\noindent
{\bf 3.} Consider three-flavor QCD in the presence of strong CP violation,
parameterized by the constant $\theta_0$,
\begin{equation}
 \mathcal{L}_{QCD} = -\frac{1}{4} G_{\mu\nu}^{a} G^{a, \mu\nu} + 
\bar q \left( i \slashed{D} - \mathcal{M} \right) q + \theta_0 
\frac{g^2}{32\pi^2} G_{\mu\nu}^{a} \tilde G^{a, \mu\nu}  \,\,\, (a = 1,
\ldots , 8) ~ ,
\label{eq:QCDLagrangian}
\end{equation}
with the gluon field strength tensor $G_{\mu\nu}^a$ and its dual
$\tilde G_{\mu\nu}^a = \varepsilon_{\mu\nu\lambda\sigma} G^{a, \lambda\sigma}$, 
$q$ collects the various quarks, and $D_\mu$ is the
gauge-covariant derivative. The last term leads to the $U(1)_A$ anomaly
and is responsible for the non-vanishing mass of the $\eta'$ in the
chiral limit. We  want to analyze the effects of strong CP violation
in the appropriate effective field theory, which is chiral perturbation 
theory. To this end, we treat the vacuum angle $\theta_0$ as an
external field and use the appropriate effective Lagrangian
for the $U(3)_L \times U(3)_R$ theory. The original,
systematic construction of the meson Lagrangian for this symmetry 
can be found in~\cite{Gasser:1984gg}. However, for our purposes it is more 
convenient to adapt the notation given in~\cite{Borasoy:2000pq}, which is 
particularly suited for the calculation of the electric dipole form factor of 
the neutron. This formulation itself is partially based 
on~\cite{Leutwyler:1996sa,HerreraSiklody:1996pm}.
As a basic building block, one
introduces the external field $\theta(x)$ that
transforms as $\theta(x) \to \theta(x) - 2N_f \alpha$ under isosinglet axial rotations, with $N_f$ the number of
active quark flavors. The mesons (the eight Goldstone bosons and the singlet
field, the $\eta_0$) are incorporated in a $3\times 3$
matrix-valued field $\tilde U$ that transforms as $\tilde U \to L\tilde U R^\dagger$
under   $U(3)_L \times U(3)_R$. Since the phase of the determinant of 
$\tilde U$ transforms according to $\ln \det \tilde U \to \ln \det \tilde U + 2 
i N_f \alpha \,$, one introduces the invariant combination
\begin{equation}
 \bar\theta = \theta - i \ln \det \tilde U \ ,
\end{equation}
which is more convenient for the construction of the effective Lagrangian.
The most general effective meson Lagrangian to second chiral order, 
complying with $U(3)_L \times U(3)_R$ symmetry, then reads
\begin{eqnarray}
 \mathcal{L} &=& - V_0 + V_1 \,\trace \bigl[ \nabla_\mu \tilde U^\dag \nabla^\mu \tilde U \bigr] 
+ V_2 \,\trace \bigl[ \tilde\chi^\dag \tilde U + \tilde\chi \tilde U^\dag \bigr] 
+ i V_3 \,\trace \bigl[ \tilde\chi^\dag \tilde U - \tilde\chi \tilde U^\dag \bigr] \nonumber \\
&+& V_4 \,\trace \bigl[ \tilde U \nabla_\mu \tilde U^\dag \bigr] \trace \bigl[ \tilde U^\dag \nabla^\mu \tilde U \bigr] 
+ V_5 \,\trace \bigl[ \nabla_\mu \theta \nabla^\mu \theta  \bigr] \ ,
\label{eq:MesonicLagrangianVEVNotFixed}
\end{eqnarray}
with  $\tilde\chi = 2 B_0 \left( s + ip \right)$ and  $\nabla_\mu
\tilde U = \partial_\mu \tilde U - i r_\mu \tilde U + i \tilde U
l_\mu$, $s$, $p$, $l_\mu$, $r_\mu$ are the standard external sources~\cite{Gasser:1984gg}.
The $V_i$ are functions of $\bar\theta$. To further analyze
these, one makes use of the large-$N_c$ approximation to QCD. 
The use of an expansion in powers of $1/N_c$ in addition to the 
usual expansion in powers of small momenta and quark masses implies an
extension  of the  power counting scheme. It is convenient to choose the 
unified counting rules according to~\cite{Leutwyler:1996sa,Kaiser:2000gs}
\begin{equation}\label{eq:count}
 p = \Order \left(\delta\right) \ , \qquad m_{q} =
 \Order\left(\delta^2\right) 
\ , \qquad 1/N_c = \Order\left(\delta^2\right) \ ,
\end{equation}
with the pertinent small parameter $\delta$, 
and $m_q$ denotes any of the light quark masses. This allows to
expand the $V_i$ in powers of $\bar\theta$, with an odd function $V_3$ 
in $\bar\theta$ and all others even; 
the expansion coefficient of $V_i$ of order $n$ will be denoted by $V_i^{(n)}$.
To utilize the effective Lagrangian
equation~\eqref{eq:MesonicLagrangianVEVNotFixed}, one must fix the vacuum
expectation value  $U_0$  of $\tilde U$ by solving the classical equations of motion. 
$U_0$ can be chosen diagonal, expressed in terms of the so-called quark angles.
This allows us to write $\tilde U = \sqrt{U_0} U \sqrt{U_0}$,
and choose the parameterization
\begin{equation}
 U = \exp \bigg(  \sqrt{\frac{2}{3}} \frac{i}{F_0} \eta_0 +  \frac{2 i }{F_{\phi}} \phi\bigg) \ .
\end{equation}
The coupling constant $F_0$ for the $\eta_0$-singlet is in principle
different from its  octet counterpart $F_{\phi}$ due to the fact that the 
unbroken subgroup $U(3)_V$ does not exhibit an irreducible
representation  of dimension nine. 
The resulting effective Lagrangian reads
\begin{eqnarray}
\mathcal{L}_{\phi} &=& - V_0 + V_1 \trace \left[ \nabla_\mu U^\dag 
\nabla^\mu U \right] + \left( V_{2} + \mathcal{B} V_3 \right) \trace 
\left[ \chi \left( U + U^\dag \right) \right] - i\mathcal{A} V_{2} \trace 
\left[ U - U^\dag\right] \nonumber \\ 
&+& i\left( V_3 - \mathcal{B} V_{2} \right) \trace \left[ \chi \left( U 
- U^\dag \right) \right] + \mathcal{A} V_3 \trace \left[ U + U^\dag 
\right] + V_4 \trace \left[ U \nabla_\mu U^\dag \right] \trace \left[ 
U^\dag \nabla^\mu U \right] \ ,
\label{eq:MesonLagrangian}
\end{eqnarray}
where the hermitian matrix $\chi$ has absorbed further factors of $U_0$, 
and $\mathcal{A}$, $\mathcal{B}$ are complicated functions
of the $V_i$, see e.g.~\cite{Borasoy:2000pq}.
In leading approximation, they are given by
\begin{equation}
\mathcal{A} = \frac{V_0^{(2)}}{V_2^{(0)}} \,\bar\theta_0 + \Order\big(\delta^4\big) ~, \quad
\mathcal{B} = \frac{V_3^{(1)}}{V_2^{(0)}} \,\bar\theta_0 + \Order\big(\delta^6\big) ~.
\end{equation}
Note that $\bar\theta_0=\Order(\delta^2)$, see below.
After the vacuum alignment and with our particular choice of
$U$, the $V_{i}$ are now functions of the combination $\bar\theta_0 
+ \sqrt{6}\eta_0 / F_0 $. Furthermore, the correct normalization 
for the terms quadratic in the singlet field is obtained by considering all 
contributions to its kinetic energy from the $V_1$- and $V_4$-terms and
demanding the resulting coefficient to equal $1/2$. Finally, we can
express $\bar \theta$ in terms of measurable quantities. For $m_{u,d} \ll m_s$, 
to lowest order in the quark angles, and neglecting numerically small corrections 
to the value of $V_0^{(2)}$~\cite{HerreraSiklody:1997kd}
this relation reads (in what follows, we set $F_\phi = F_\pi$)
\begin{equation}
 \bar\theta_0 = \frac{F_\pi^2 M_\pi^2}{8 V_0^{(2)}} \theta_0 \ .
\label{eq:theta0barMpi}
\end{equation}

In a similar manner, one can construct the most general effective Lagrangian
including also the baryon octet $B$  up-to-and-including terms of second order 
in the derivative expansion (we only display the terms relevant to our
calculation; for details, we refer to~\cite{Borasoy:2000pq}):
\begin{eqnarray}
 \mathcal{L}_{\phi B} &=& i \,\trace \bigl[ \bar{B} \gamma^\mu [ D_\mu , B ] \bigr] 
- \chiral{m} \,\trace [ \bar{B} B ] 
- \frac{D/F}{2} \trace \bigl[ \bar{B} \gamma^\mu \gamma_5 [ u_\mu, B ]_\pm \bigr] 
- \frac{\lambda}{2} \,\trace \bigl[ \bar{B} \gamma^\mu \gamma_5 B \bigr] \trace [ u_\mu ] \nonumber \\
 &+& b_{D/F} \,\trace \bigl[ \bar{B} \big[ \chi_+ - i \mathcal{A} (U-U^\dag), B \big]_\pm \bigr] 
+ b_0 \,\trace [ \bar{B} B ] \,\trace \big[ \chi_+ - i \mathcal{A} (U-U^\dag)\big]  
\nonumber\\
 &+& 4 \mathcal{A} \,w'_{10} \frac{\sqrt{6}}{F_0} \eta_0 \trace [ \bar{B} B ]
+ i \Big( w'_{13/14} \,\bar\theta_0 + w_{13/14} \frac{\sqrt{6}}{F_0} \eta_0 \Big) 
\trace \bigl[ \bar{B} \sigma^{\mu\nu} \gamma_5 [ F_{\mu\nu}^{+} , B ]_\pm \bigr] \nonumber \\
&+& w_{16/17}\, \trace \bigl[ \bar{B} \sigma^{\mu\nu} [ F_{\mu\nu}^{+}, B ]_\pm \bigr] \ , 
 \label{eq:BaryonLagrangian}
\end{eqnarray}
utilizing  $U = u^2$,
$\Gamma_\mu = \big[ u^\dag\left(\partial_\mu - i r_\mu\right) u  
                  + u     \left(\partial_\mu - i l_\mu\right) u^\dag\big]/2$, and 
$ [D_\mu, B] = \partial_\mu B + [\Gamma_\mu, B]$.
Furthermore,
$u_\mu = i\big[ u^\dag \left(\partial_\mu - i r_\mu\right) u
              - u      \left(\partial_\mu - i l_\mu\right) u^\dag\big]$,
$\chi_+ = u\chi^\dag u+u^\dag\chi u^\dag$, and $F_{\mu\nu}^+$
incorporates the field strength tensor.
Also, $\chiral{m}$ is the octet mass in the chiral limit, $F$, $D$ are the
conventional axial coupling constants,  $\lambda$ is an isosinglet axial
coupling, and the $b_{0/D/F}$ are the low-energy constants (LECs) related to the leading-order
symmetry breaking. The LECs $w_i$ parameterize the coupling of the singlet
field to the baryons at second order (we use the abbreviation $w'_{10} = w_{10}+3w_{12}/2$
compared to~\cite{Borasoy:2000pq}), except for $w_{16/17}$ which are the conventional
magnetic moment couplings.

\begin{figure}
\begin{center}
\includegraphics[width=35mm]{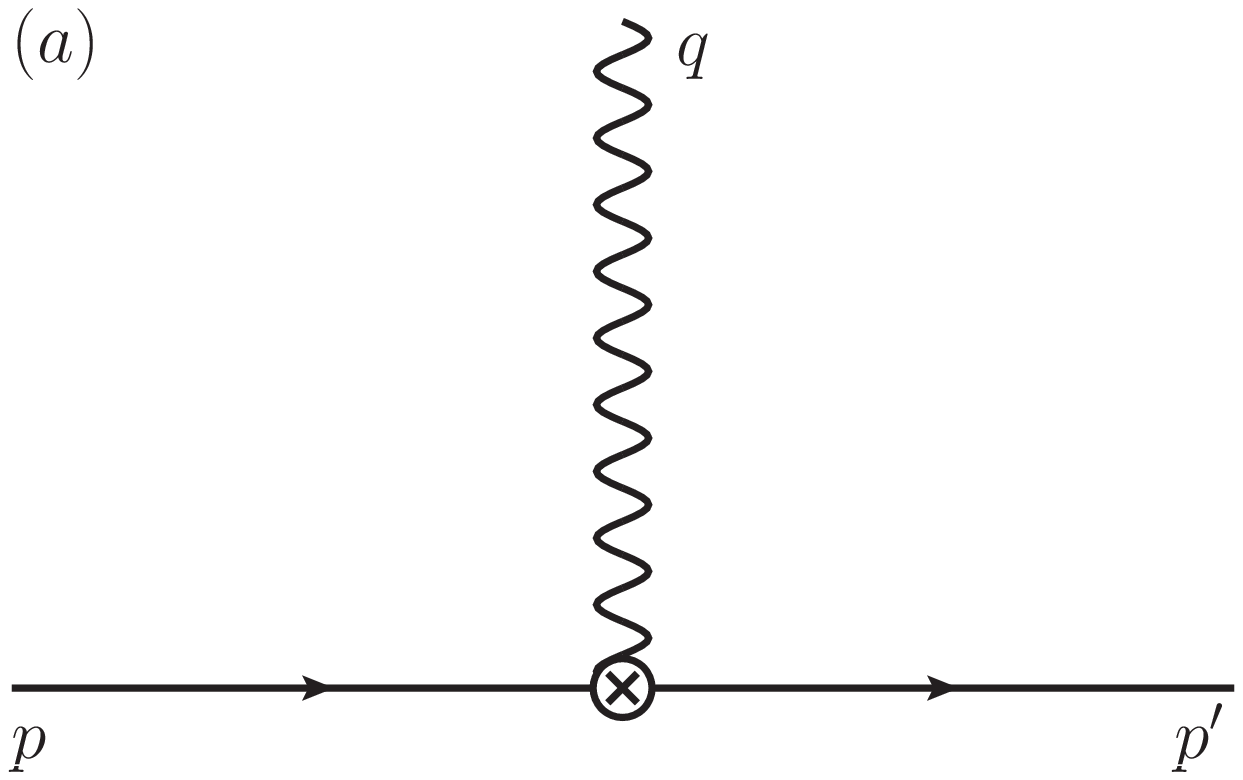}
\includegraphics[width=35mm]{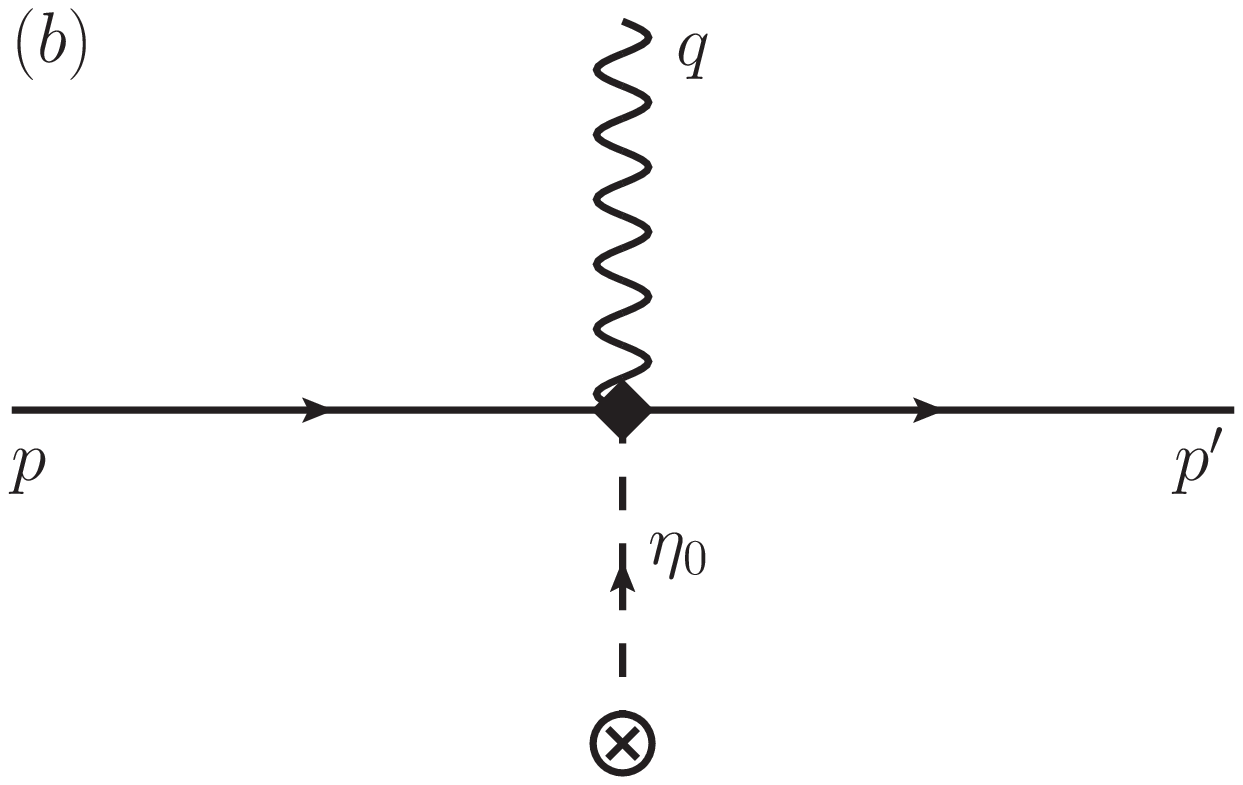}
\includegraphics[width=35mm]{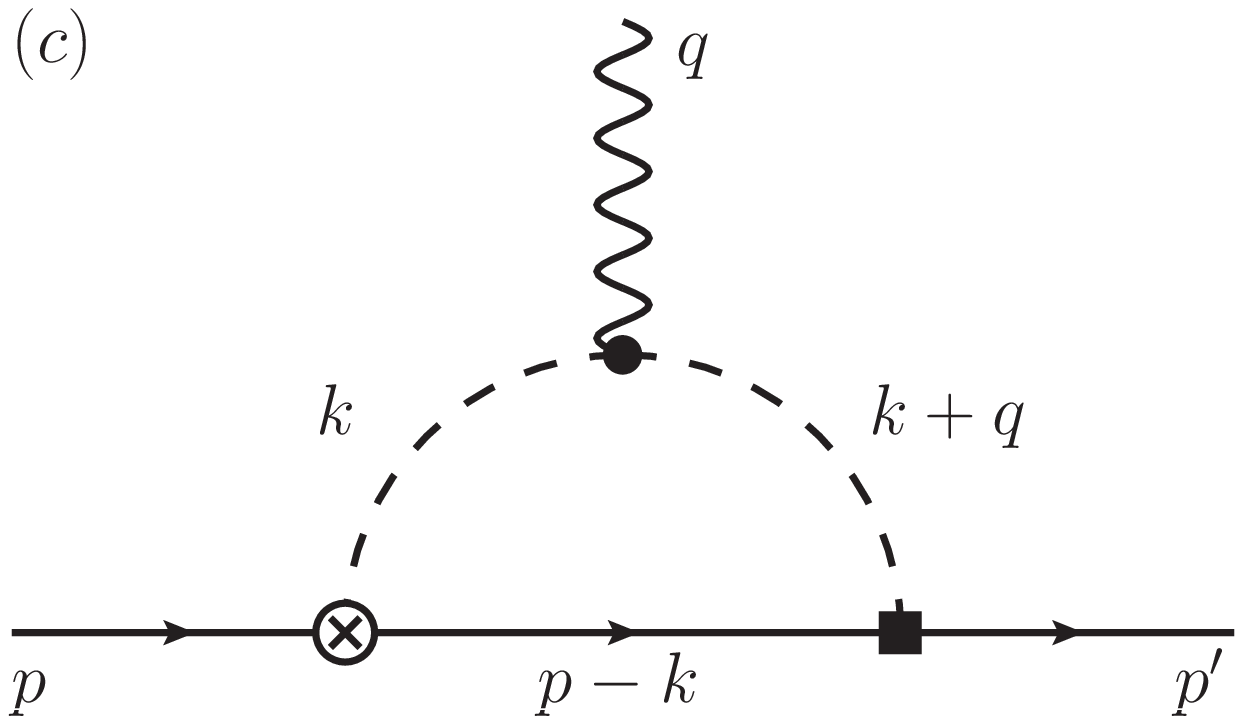}
\includegraphics[width=35mm]{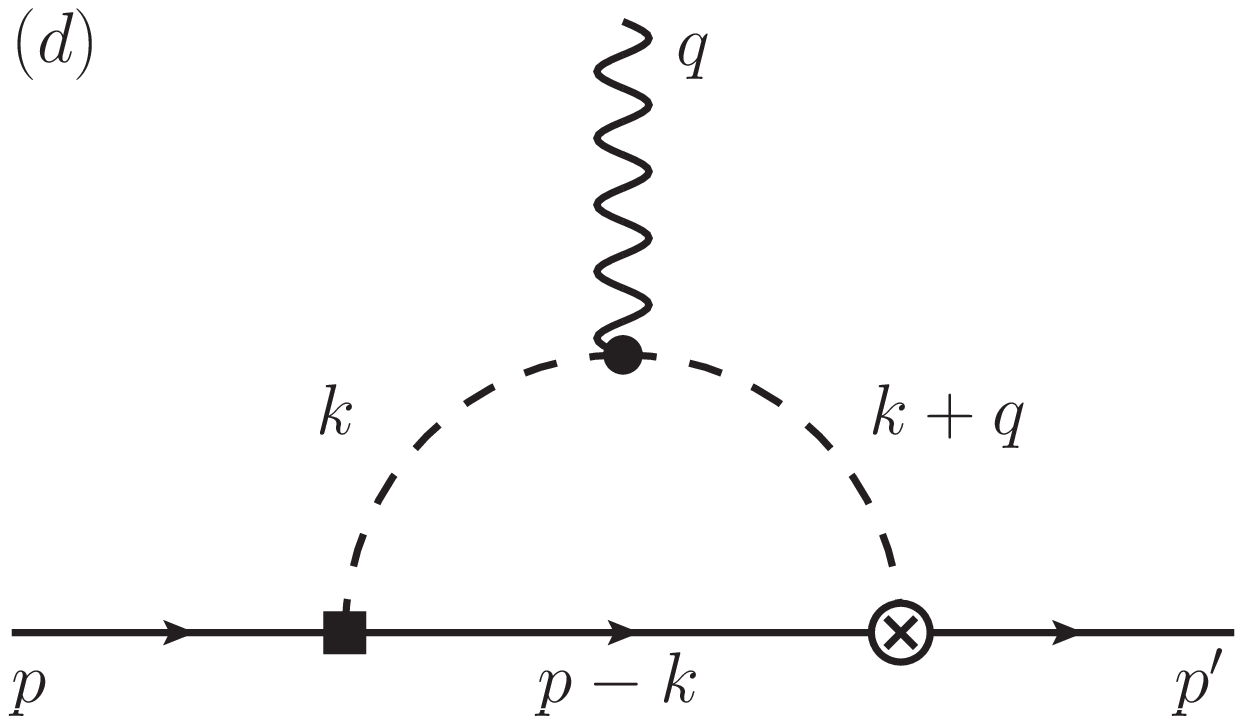}

\vspace{5mm}

\includegraphics[width=35mm]{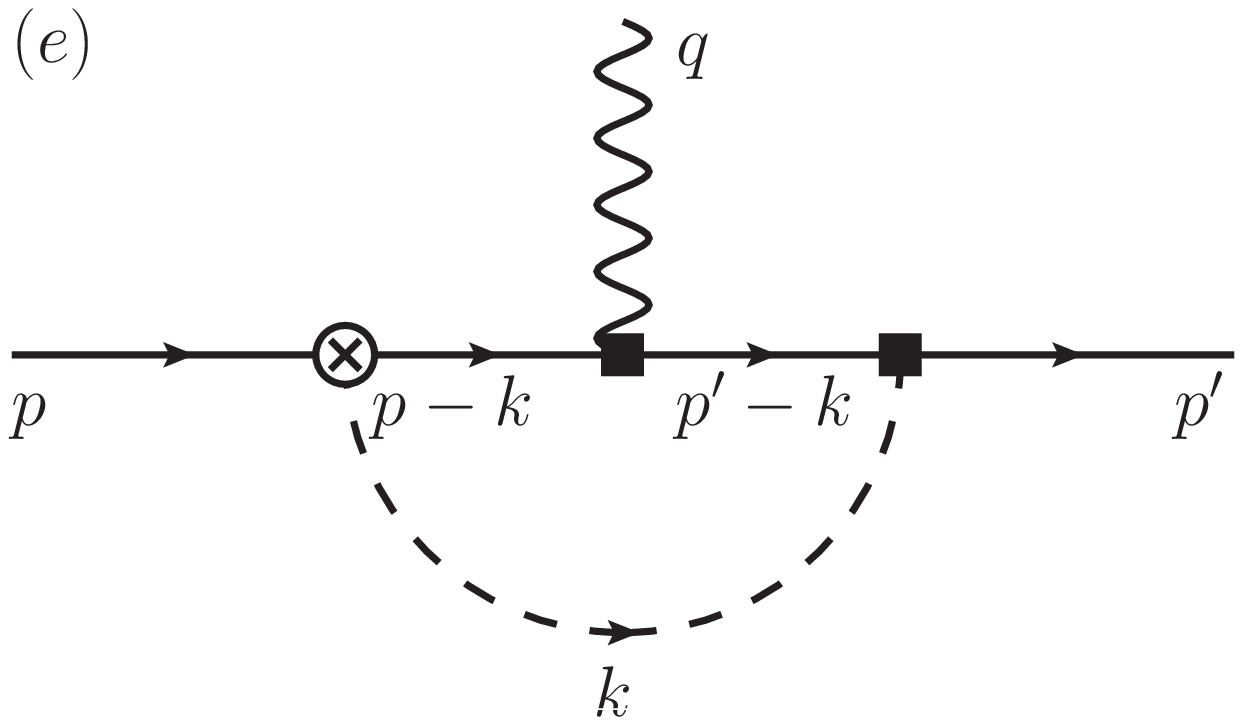}
\includegraphics[width=35mm]{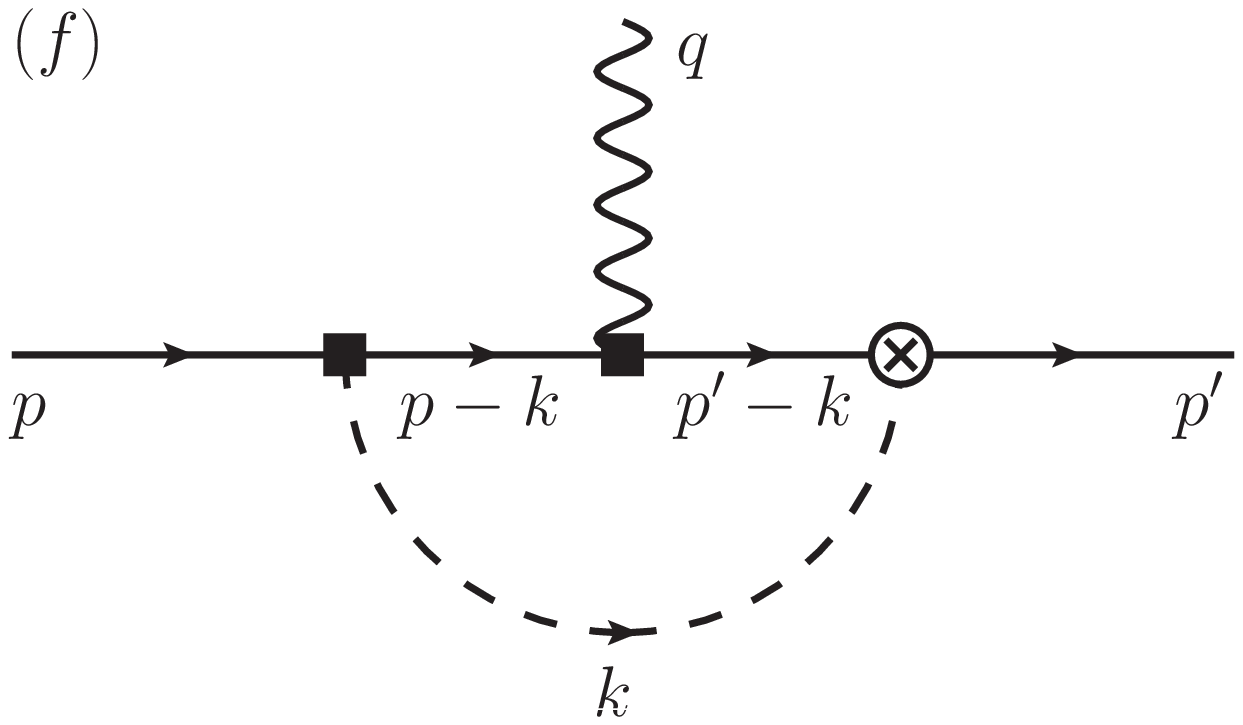}
\includegraphics[width=35mm]{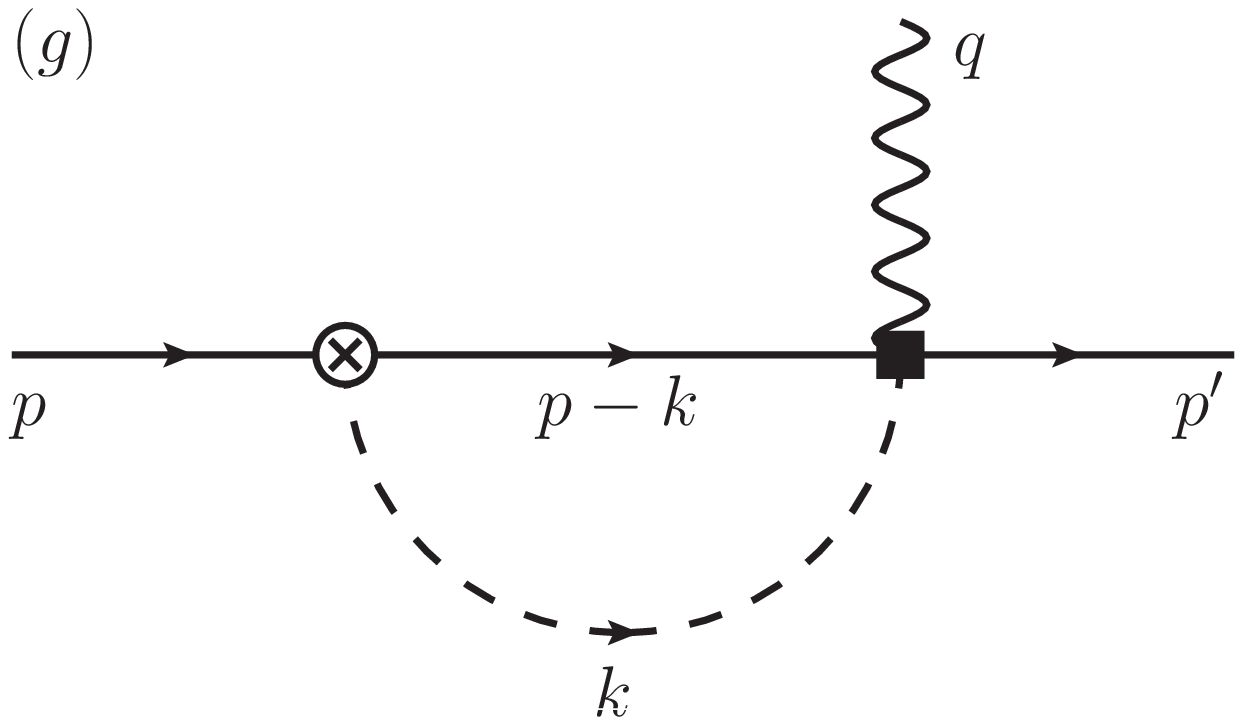}
\includegraphics[width=35mm]{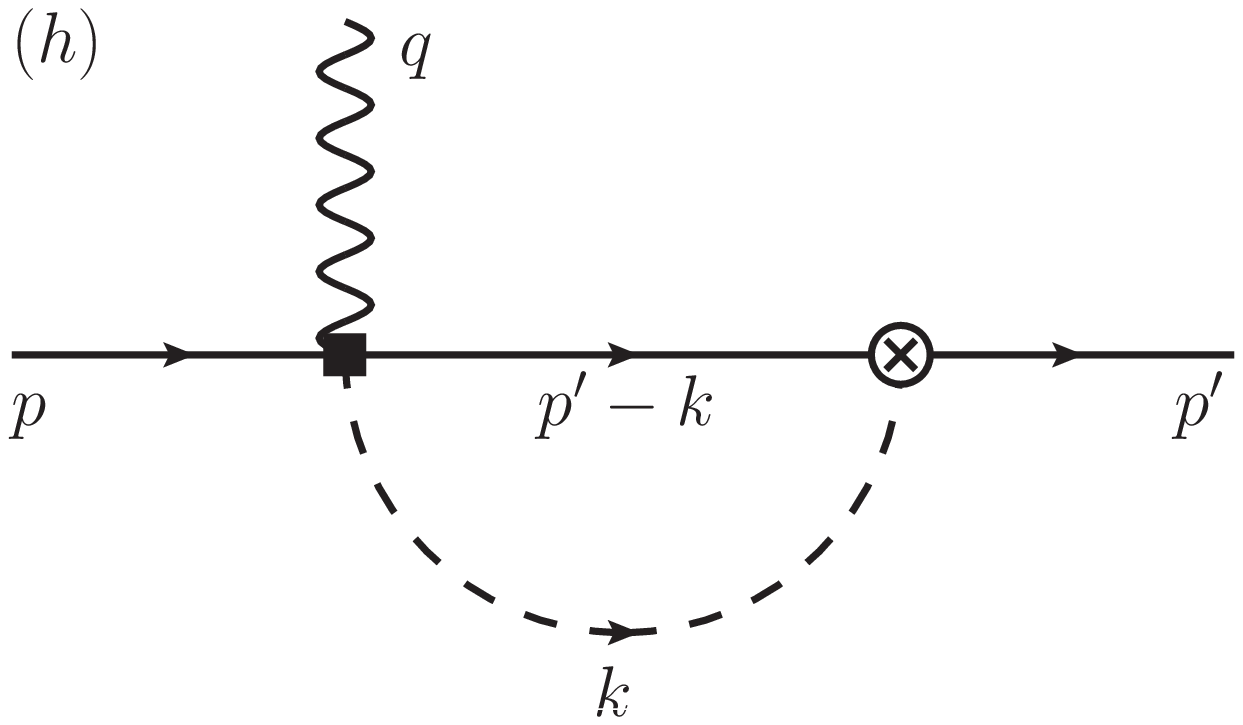}

\vspace{5mm}

\includegraphics[width=35mm]{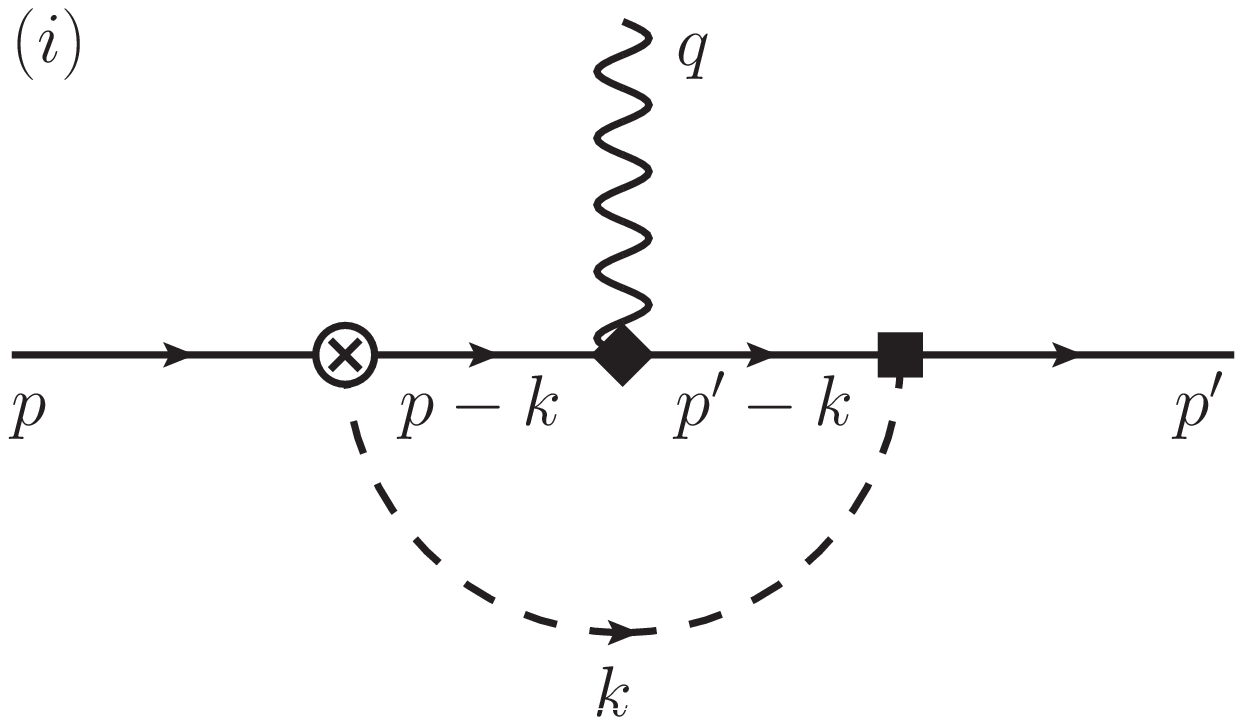}
\includegraphics[width=35mm]{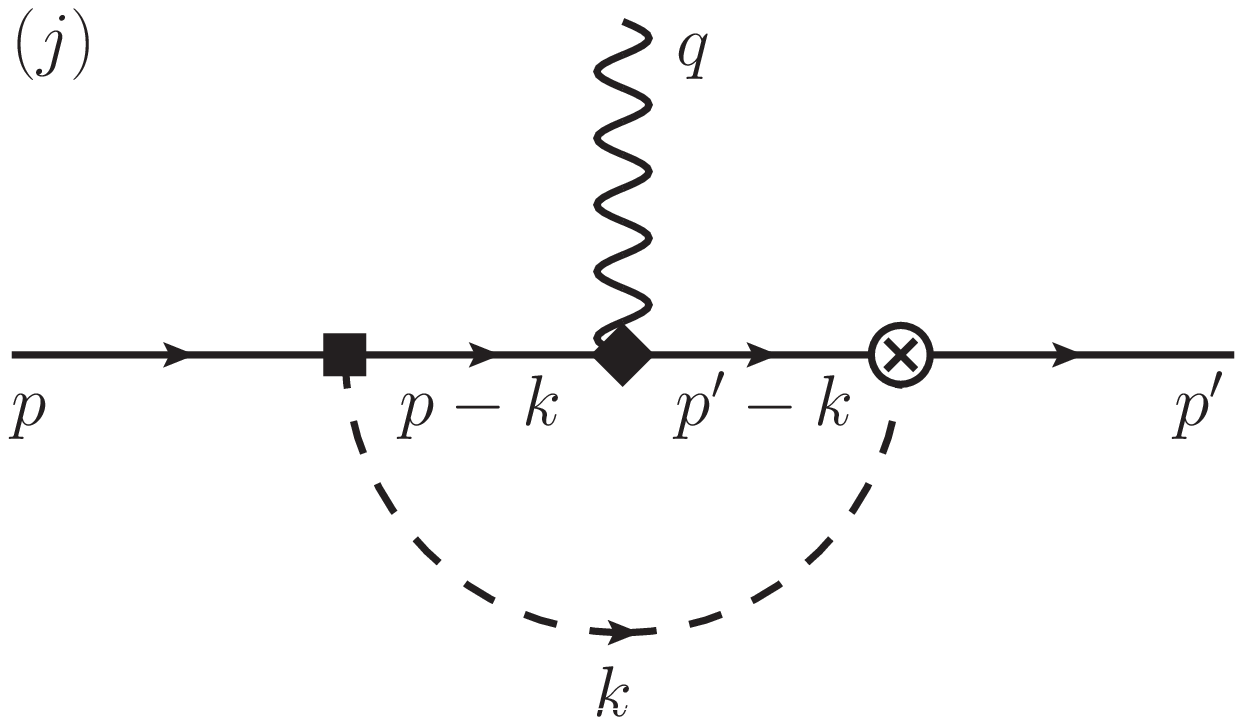}
\includegraphics[width=35mm]{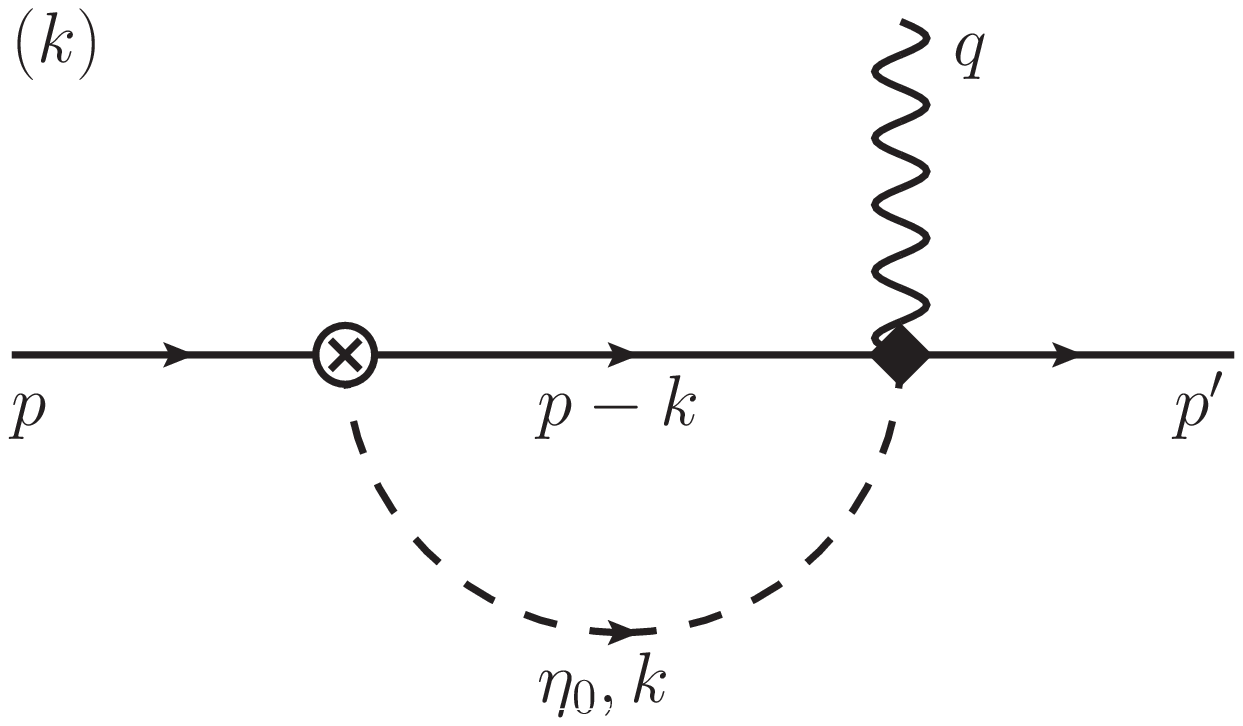}
\includegraphics[width=35mm]{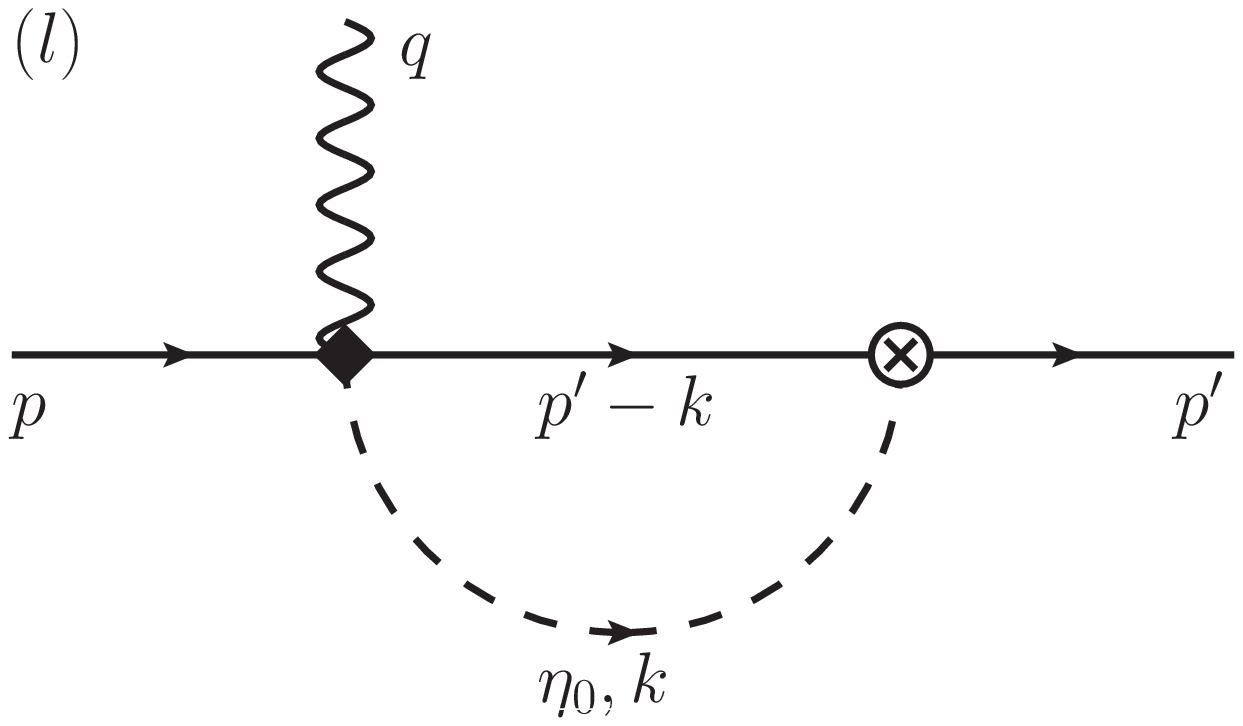}

\caption{Feynman diagrams contributing to the electric dipole form 
factor of the nucleon. Diagrams (a), (b) contribute at tree level, (c)--(h) 
are of the second and (i)--(l) of the third chiral order. Filled circles are 
second-order mesonic vertices, squares and diamonds represent vertices 
generated by the first- and second-order meson--baryon Lagrangian,
respectively. CP-violating vertices are denoted by $\otimes$. 
}
\label{fig:dia}
\end{center}
\end{figure}

\medskip
\noindent
{\bf 4.} We now have assembled all pieces to calculate the dipole
form factor of the neutron (and the proton) to third order in the
chiral expansion. In Fig.~\ref{fig:dia} we show the corresponding
Feynman graphs. 
\noindent These are tree graphs (a), (b), one-loop graphs starting at 
$\Order(\delta^2)$ (c)--(h), and one-loop graphs with one insertion from the
dimension-two chiral meson--baryon Lagrangian starting at $\Order(\delta^3)$ 
(i)--(l). To obtain a consistent power counting, the loop diagrams
are evaluated using the so-called infrared regularization 
scheme~\cite{Becher:1999he}. In this covariant framework, 
loop graphs resum an infinite string of higher-order corrections,
associated with the $1/m$ insertions in the heavy-baryon propagator.
In the following, we will expand these contributions
up-to-and-including $\Order(\delta^3)$. It can be shown that the graphs
(g)--(l) do not contribute at this order or cancel each other exactly. 
For later purpose, it is convenient to decompose the form factor into tree level 
and loop contributions, $F_3  = F_3^{\rm tree} + F_3^{\rm loop}\,$. After renormalization,
the tree level part for the neutron  is given by
\begin{equation}
 F_{3,n}^{\rm tree} = -16 \, m \, e \, \bar\theta_0 \left[ \frac{1}{3} w_{13}^r 
+ \frac{48}{F_0^2 F_\pi^2M_{\eta_0}^2} V_3^{(1)} 
V_0^{(2)} w_{13} \right] \ ,
 \label{eq:nEDFF_tree}
\end{equation}
where $w_{13}^r$ is the finite part of $w'_{13}$.
The third-order loop result for the neutron electric dipole form factor
follows as
\begin{eqnarray}
 F_{3,n}^{\rm loop}\left( q^2 \right) 
&=& \frac{2 m e V_0^{(2)} \bar\theta_0}{\pi^2 F_\pi^4}  
\Biggl\{ (D+F) (b_D+b_F) \biggl[ 1 - \ln \frac{M_\pi^2}{\mu^2} - \sigma_\pi
\ln\frac{\sigma_\pi + 1}{\sigma_\pi - 1} 
- \frac{\pi}{2} \frac{q^2 - 2 M_\pi^2}{m \sqrt{-q^2}} \arctan \frac{\sqrt{-q^2}}{2 M_\pi}\biggr]
\nonumber \\
-  (D&-&F)(b_D-b_F) \biggl[ 1 -\ln \frac{M_K^2}{\mu^2}
- \sigma_K \ln \frac{\sigma_K + 1}{\sigma_K - 1} 
- \frac{\pi}{2} \frac{ q^2 - 2 M_K^2 + 4 m \Delta m_{\Sigma N}}{m \sqrt{-q^2}} 
\arctan \frac{\sqrt{-q^2}}{2 M_K}\biggr] \Biggr\} \,, \nonumber\\ && 
\label{eq:nEDFF_O3}
\end{eqnarray}
where, in the space-like region, $q^2=-Q^2 \leq 0$, $\Delta m_{\Sigma N} = m_\Sigma - m$,
which is of order $\delta^2$, and
$\sigma_{\pi/K} = \sqrt{1 - 4 M_{\pi/K}^2/q^2}$.
The second-order result is given by omitting the last term of the pion and
kaon part in the expression above. In the limit $q^2 \to 0$, one can readily
deduce the expression for the electric dipole moment of the neutron,
\begin{eqnarray}
d_n &=& d_n^{\rm tree} +  d_n^{\rm loop}~, \qquad 
d_n^{\rm tree} = -8 \, e\, \bar\theta_0 \left( \frac{1}{3} w_{13}^r 
+ \frac{48}{F_0^2 F_\pi^2M_{\eta_0}^2} V_3^{(1)} 
V_0^{\left( 2 \right)} w_{13} \right) ~ ,  \nonumber\\
 d_n^{\rm loop} &=& - \frac{e \, V_0^{(2)} }{\pi^2F_\pi^4} 
\bar\theta_0 \Biggl[  (D+F) (b_D+b_F) 
\left( 1 + \ln{ \frac{M_\pi^2}{\mu^2}} - \frac{\pi M_\pi}{2 m} \right)  \nonumber\\
&& \qquad\qquad\,\, - (D-F) (b_D-b_F) \left( 1 
+ \ln \frac{M_K^2}{\mu^2} - \frac{\pi M_K}{2 m} 
+ \frac{\pi \,\Delta m_{\Sigma N}}{M_K}\right) \Biggr] \ .
\end{eqnarray}
The expression for the tree result was already  derived in
\cite{Borasoy:2000pq}, however we obtain an additional factor of three for 
diagram (b). 
Note that we have kept the dependence on the scale of
dimensional regularization $\mu$ in the loop expression in order to give an 
error estimate by varying the scale when performing the numerical evaluation.
Dropping all  terms besides the chiral logarithms reproduces the result 
of~\cite{Pich:1991fq,Borasoy:2000pq}. 
The next-to-leading order correction in the pion-loop contribution was also
obtained in~\cite{Narison:2008jp} in a relativistic loop calculation.
We have also calculated the 
corresponding form factor of the proton to achieve an isoscalar/isovector
separation in the two-flavor case as discussed below. We refrain from
giving the corresponding formulas here~\cite{KO}. From the neutron and
proton form factors one can derive analytic results for the corresponding
radii, the one for the neutron reads
\begin{eqnarray}
 \left\langle r_{ed}^2 \right\rangle_{n} = \frac{2 e
   V_0^{(2)}}{\pi^2 F_\pi^4} \bar\theta_0 \Biggl[ 
 (D&+&F) (b_D+b_F) \left( \frac{m}{M_\pi^2} - \frac{5\pi}{4 M_\pi} \right)  
\nonumber \\ 
- (D &-& F) (b_D-b_F) \left( \frac{m}{M_K^2} -  \frac{5\pi}{4 M_K} - \frac{\pi \,m\, 
\Delta m_{\Sigma N}}{2 M_K^3} \right) \Biggr]  ~. \label{eq:nEDMSR_O3}
\end{eqnarray}

\medskip
\noindent
{\bf 5.} Before showing results, we must fix parameters. 
We use $D = 0.804$ and $F=0.463$ from hyperon $\beta$-decays~\cite{Abele:2003pz}, 
$b_D = 0.066\,{\rm GeV}^{-1}$ and 
$b_F = -0.209\,{\rm GeV}^{-1}$ from the leading-order analysis of the baryon mass splittings, $F_\pi =
92.4\,$MeV as well as $V_0^{(2)} = -5\cdot 10^{-4}\,$GeV$^4$ and 
$V_3^{(1)} = 3.5\cdot 10^{-4}\,$GeV$^2$ from an analysis of $\eta-\eta'$ mixing~\cite{HerreraSiklody:1997kd}.
The same analysis also shows that the correction to  $F_0/F_\pi = 1$ is of
order $1/N_c$, so that we use $F_0 = F_\pi$ based on the counting rule in Eq.~\eqref{eq:count}.

First, we consider the neutron dipole moment. Following the arguments
of~\cite{Borasoy:2000pq} based on large-$N_c$ counting, the ratio of
diagrams (a) and (b) is given by 
\begin{equation} 
 \Bigg| \frac{d_n^{\rm tree(b)}}{d_n^{\rm tree(a)}} \Bigg| 
\simeq \frac{144}{F_\pi^4 M_{\eta_0}^2} \Big| V_0^{(2)} V_3^{(1)} \Big| \simeq 0.38 \ .
\end{equation}
This ratio will serve for an error estimate of the tree level contribution.
Furthermore, the $w_{16}$-term of the meson--baryon Lagrangian, describing a magnetic coupling
to the baryon fields, is similar to the structure proportional to $w'_{13}$ / $w_{13}^r$
except for the additional $\gamma_5$. 
However, the coefficient $w_{16}$ is of $\Order(N_c^1)$, whereas $w_{13}^r$ is of 
$\Order(N_c^0)$, which suggests the assumption
$| w_{13}^r | < | w_{16} |$.  Utilizing the third-order calculation of the 
baryon electromagnetic form factors~\cite{Kubis:2000aa}, 
we find $w_{16} = 0.40\,{\rm GeV}^{-1}$. Assuming further  $|w_{13}^r| = |w_{16}| / 3$, 
we obtain a bound on the dipole moment,
 \begin{equation}
\left| d_n^{\rm tree} \right| \simeq \left( 2.9 \pm 1.1 \right) 
\times  10^{-16} \, \theta_0 \, e \,\mbox{cm} \ .
\label{eq:nEDMTreeResult2}
\end{equation}
The numerical evaluation of the loop corrections to the electric dipole moment
of the neutron  is much more straightforward since it
does not involve unknown parameters. We find
\begin{equation}
 d_n^{\rm loop} = - 3.0_{-0.8}^{+1.1} \times 10^{-16} \, \theta_0 \, e \, \mbox{cm}  \ ,
\label{eq:nEDMloopresultNLO}
\end{equation}
varying the renormalization scale $\mu$ between $M_\rho$ and $m_\Xi$ (note the much larger
scale variation in~\cite{Narison:2008jp} that induces an even more pronounced
uncertainty).  However, it was shown 
in~\cite{Borasoy:1996bx} that the corrections to $b_D$, $b_F$ due to
fourth-order 
effects are large, so that the central value for the loop contributions changes to 
$
d_n^{\rm loop} =  - 5.1 \times 10^{-16} \, \theta_0 \, e \, \mbox{cm} .
$
Taking this latter value, the loop corrections slightly dominate and we can deduce a 
lower bound for the theoretical  estimate of the electric dipole moment, 
\begin{equation}
 \left| d_n^{\rm theo} \right| \gtrsim  1.1 \times 10^{-16} \, \theta_0 \, e \, \mbox{cm} \ .
\end{equation}
 Together with the experimental upper bound on the electric dipole moment of
 the neutron, this finally yields
\begin{equation}\label{eq:bound}
 \left| \theta_0 \right| \lesssim 2.5 \times 10^{-10} \ .
\end{equation}
We remark, however, that this bound is very sensitive to cancellations between the tree and loop contributions
and based on large-$N_c$ arguments, and thus should be considered with caution. 

\begin{figure}
\begin{center}
\includegraphics[width=83mm]{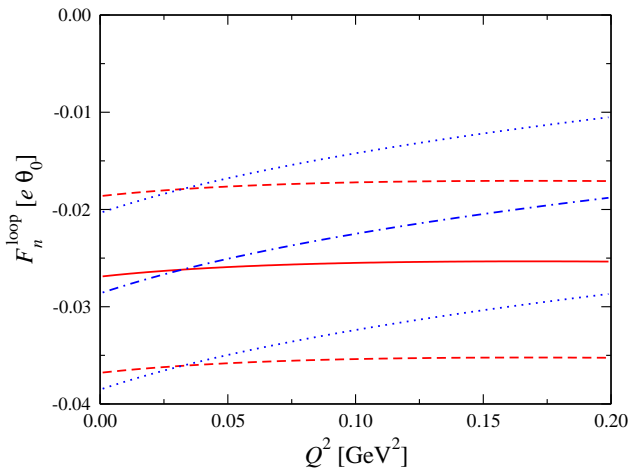}
\caption{Loop contribution to the electric dipole form factor of the neutron
  as a function of  $Q^2=-q^2$ to third (red, solid line) and second (blue,
  dot-dashed line) order, using $\mu = 1 \, \mbox{GeV}$. The boundaries of the 
  error bands due to variation of the renormalization scale $\mu$ between 
  $M_{\rho}$ and $m_{\Xi}$ are represented by red dashed or blue dotted lines, respectively.}
\label{fig:ff}
\end{center}
\end{figure}

Next, we discuss the loop contribution to the neutron dipole form factor. 
It is shown in Fig.~\ref{fig:ff} in units of $e$ times the 
dimensionless  parameter $\theta_0$. 
The full result including third-order corrections is represented
by  the solid line, whereas the dashed line contains second-order 
contributions only. The error bands of both lines are obtained by varying the 
renormalization scale between $M_\rho$ and $m_\Xi$ again. 
The renormalization scale for the central curves has been set to the 
neutron mass $\mu = m$. As can be seen from this figure, the relative size of 
the third-order corrections compared to the leading-order terms is small for 
reasonable values of $Q^2$. In fact, the change due to the variation of the 
renormalization scale clearly exceeds the change resulting from third-order 
corrections. Nevertheless, the slope of the form factor for small values of 
$Q^2$ is altered significantly by the next-to-leading-order corrections, as
it is also borne out by evaluating Eq.~\eqref{eq:nEDMSR_O3}. We find
\begin{equation}
\langle r^2_{ed}\rangle_n = -20.4 \, \left[1 - 0.67 + \Order(\delta^2)\right]\,
\theta_0 \, e \, {\rm fm}^2~,
\end{equation}
and a similar effect is found for the proton. This large correction can be traced back to the
additional factors of $\pi$ in the pion loop contribution in
Eq.~\eqref{eq:nEDMSR_O3}. Such factors also appear in the analysis of the
radii of the
isospin-violating nucleon form factors, see~\cite{Kubis:2006cy}. We also note that the contribution from
kaon loops is much smaller than the one from the pions and therefore, the radius
is almost entirely of isovector nature.

\begin{figure}
\begin{center}
\includegraphics[width=83mm]{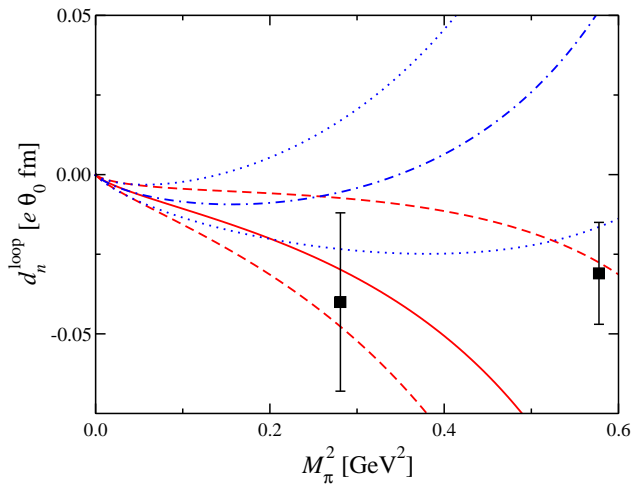}
\caption{
  Plot of $d_n^{\rm loop}$ as a function of the pion mass $M_\pi^2$ to third (red,
  solid line) and second (blue, dot-dashed line) order, with $\mu = 1 \,
  \mbox{GeV}$.  The boundaries of the error bands due to variation of the 
  renormalization scale $\mu$ between $M_{\rho}$ and $m_{\Xi}$ are represented 
  by red dashed and blue dotted lines, respectively. The data from two-flavor lattice
  QCD \protect\cite{Shintani:2008nt} are shown by the black squares.}
\label{fig:massdep}
\end{center}
\end{figure}

To compare results from two-flavor lattice QCD at unphysical quark
masses with predictions from chiral perturbation theory, it is necessary to
perform an extrapolation of the analytic results in the pion mass. 
To this end, we have to express $\Delta m_{\Sigma N}$ in terms
of $M_\pi$, $M_K$ and account for the fact that $M_K$ itself depends on $M_\pi$. 
To lowest order this dependence is given by
\begin{equation}
 M_K^2 = \chiral{M}_K^2 + \frac{1}{2} M_\pi^2 \ ,
\label{eq:KaonMass1}
\end{equation}
where $\chiral{M}_K$ denotes the mass of the kaon in the
$SU(2)$ chiral limit. Numerically, one has $ \chiral{M}_K \simeq 0.484 \, \mbox{GeV}  .$
Special care has to be taken furthermore 
to relate $\bar\theta_0$ to $\theta_0$, as Eq.~\eqref{eq:theta0barMpi} 
has been derived assuming $m_{u,d} \ll m_s$, as well as a certain hierarchy
of terms relying on the numerical value of $V_0^{(2)}$.
Giving up those assumptions changes the relation between $\bar\theta_0$ and $\theta_0$ to
\begin{equation}
 \bar\theta_0 = \left( 1 + \frac{4 V_0^{(2)}}{F_\pi^2} 
\frac{4 M_K^2 - M_\pi^2}{M_\pi^2\left(2 M_K^2 - M_\pi^2 \right)}\right)^{-1}  \,\theta_0 \ .
\label{eq:theta0barMK}
\end{equation}
In Fig.~\ref{fig:massdep} we show the resulting pion (quark) mass dependence
of the loop contribution to the electric dipole moment in comparison to the 
available data points from two-flavor lattice QCD~\cite{Shintani:2008nt}.
It is interesting to see that the third-order calculation reproduces the
trend of the lattice data (the order of magnitude and the global sign) even without the unknown tree contribution.
However, only below pion masses of the order of 500~MeV the third-order 
corrections are sufficiently small for a  stable chiral extrapolation.

\bigskip
\noindent
{\bf 6.} Finally we consider the matching of our three-flavor results to the
two-flavor representation. This not only allows to extend the findings
of~\cite{Hockings:2005cn} to higher order, but also gives an explicit 
representation for some of the LECs appearing in that paper. 
Such matching relations between three- and two-flavor versions of
chiral perturbation theory have been studied extensively in the meson 
sector (see the most recent comprehensive results~\cite{GHIS1,GHIS2} and references therein),
and also first results for the meson--baryon sector exist~\cite{Frink:2004ic,Mai:2009ce}.

The leading-order $SU(2)$ formulas for the electric dipole
moment of the nucleon have been worked out in~\cite{Hockings:2005cn}, 
within the framework of heavy-baryon chiral perturbation theory. The authors 
employ a decomposition of the electric dipole form factor of the nucleon 
according to 
\begin{equation}
 J_{ed}^{\nu} = - \frac{1}{2m} \sigma^{\mu\nu}q_\mu\gamma_5
\left( F_3^{(0)}(q^2)
+  F_3^{(1)}(q^2)\tau_3 \right) \ , \label{eq:EMCurrentHockings}
\end{equation}
where $F_3^{(0)}(q^2)$ is the isoscalar and $F_3^{(1)}(q^2)$ is the isovector part of the electric dipole form factor.
The relations between the corresponding isoscalar and isovector electric dipole moments $d_{0/1}$
and those of proton and neutron are given by
\begin{equation} 
 d_{0/1} = \frac{1}{2} ( d_p \pm d_n ) \ .
\label{eq:d01Def}
\end{equation}
By matching our $SU(3)$ result to the two-flavor theory, we can extend the
expressions given in~\cite{Hockings:2005cn} by one order in the chiral expansion,
\begin{equation}
d_0 = \tilde d_0 + \frac{e g_A \bar g_{\pi NN}}{8\pi^2F_\pi}\left(
  -\frac{3\pi}{4}\frac{M_\pi}{m}\right)~,\qquad
d_1 = \tilde d_1 + \frac{e g_A \bar g_{\pi NN}}{8\pi^2F_\pi}\left(
1 + \ln\frac{M_\pi^2}{\mu^2} - \frac{5\pi}{4}\frac{M_\pi}{m}\right)~, \label{eq:SU2res}
\end{equation}
where the $\tilde d_{0/1}$ parameterize the short-distance physics that now includes
kaon, $\eta_8$, and $\eta_0$ loops.  Explicitly, the matching relations for these two constants read
\begin{eqnarray}
\tilde d_{0/1} &=& \frac{1}{2}\big(\tilde d_p^{\rm tree}+\tilde d_p^{\rm loop}\big) \pm 
             \frac{1}{2}\big(\tilde d_n^{\rm tree}+\tilde d_n^{\rm loop}\big) + \Order\big(M_K^2,M_{\eta_0}^2\big)~,
\nonumber\\
\tilde d_p^{\rm tree} &=& -\frac{\tilde d_n^{\rm tree}}{2}+ 12 \,e\, \bar\theta_0 \bigg( \frac{w_{14}^r}{3} + 
\frac{48V_3^{(1)} V_0^{(2)} }{F_0^2 F_\pi^2 M_{\eta_0}^2}  w_{14} \bigg) ~,\quad
\tilde d_n^{\rm tree} = -8 \,e \,\bar\theta_0 \bigg( \frac{w_{13}^r}{3} 
+ \frac{48V_3^{(1)} V_0^{\left( 2 \right)}}{F_0^2 F_\pi^2M_{\eta_0}^2}  w_{13} \bigg) ~,  \nonumber\\
\tilde d_p^{\rm loop} &=& \frac{\tilde d_n^{\rm loop}}{2} + \frac{e \, V_0^{(2)} }{\pi^2F_\pi^{4}} \bar\theta_0 \Biggl[ 
 \frac{1}{6} (D+3F) (b_D+3b_F)  \left( 1 + \ln\frac{M_K^2}{\mu^2} - \frac{3\pi M_K}{2 m} 
- \frac{4}{3}\pi(b_D+3b_F)M_K \right)  \nonumber \\
&&-  \bigg( \frac{3}{2}(D-F) (b_D-b_F) + \frac{(D-3 F) (b_D-3b_F)}{3\sqrt{3}}\bigg) \frac{\pi M_K}{m}  \nonumber\\
&&  
- (2D+3\lambda) \left( 2 b_D + 3 b_0 + 6\sqrt{3} w'_{10} \right) \frac{\pi M_{\eta_0}}{6\sqrt{3}m}\frac{F_\pi^2}{F_0^2}  \Biggr] ~,
\nonumber \\
\tilde d_n^{\rm loop} &=& \frac{e \, V_0^{(2)} }{\pi^2F_\pi^4} 
\bar\theta_0 \Biggl[ (D-F) (b_D-b_F) \left( 1 
+ \ln \frac{M_K^2}{\mu^2} - \frac{\pi M_K}{2 m} 
+ 4\pi(b_D-b_F) M_K \right) \Biggr]  ~,
\end{eqnarray}
where we have used the $SU(2)$ chiral limit relation $M_{\eta_8}=2M_K/\sqrt{3}$.
The new terms in the $SU(2)$ expressions Eq.~\eqref{eq:SU2res} are the ones linear in the pion mass. 
We note that the corrections $\sim M_\pi$ are again sizeable.
Furthermore, it was argued in~\cite{Hockings:2005cn} based on dimensional 
analysis that the CP-violating pion--nucleon coupling $\bar g_{\pi NN}$ has to be of order\begin{equation}
 \bar g_{\pi NN}=\Order\bigg( \frac{\theta_0 M_\pi^2}{m\,F_\pi} \bigg) \ .\label{eq:Orderg0bar}
\end{equation} 
The previous results  allow us to obtain a matching relation for
$\bar g_{\pi NN}$. Utilizing the standard matching relation for the two- and
three-flavor axial couplings, $g_A = D+F + \Order(\delta^2)$, we 
obtain for $\bar g_{\pi NN}$
\begin{equation}
 \bar g_{\pi NN} = \frac{\theta_0 M_\pi^2}{F_\pi}\left( b_D 
+ b_F \right) + \Order \left( \delta^4 \right) \, ,
\label{eq:matching_g0bar}
\end{equation}
which numerically yields $\bar g_{\pi NN} = - 0.03 [-0.05] \, \theta_0$,
where the number in the brackets refers to the values of $b_{D/F}$ 
from~\cite{Borasoy:1996bx}.
This confirms the dimensional estimate equation~\eqref{eq:Orderg0bar} and
reproduces the result  originally obtained in~\cite{Crewther:1979pi}.
However, our representation in terms of symmetry-breaking LECs only is
certainly more compact and can be systematically improved by going to 
yet higher orders in the chiral expansion.

\medskip
\noindent
{\bf 7.} In this Letter, we have analyzed the neutron (and proton)
electric dipole form factor and its moment in the framework of
covariant $U(3)_L \times U(3)_R$ baryon chiral perturbation theory,
extending earlier calculations by one order in the power counting.
The latter includes the expansion of QCD in the number of colors.
While the third-order corrections to the neutron electric dipole moment are small at the physical pion mass,  
their contribution makes the trend of the quark-mass expansion agree with the
existing data from full lattice QCD, although those still employ pion masses 
too high for a safe chiral extrapolation.
More lattice data at smaller pion masses will allow one to analyze strong CP violation as encoded
in the neutron electric dipole moment in more detail. 
From comparison with experimental limits,
we have given an upper bound on the vacuum angle.
The electric dipole radius, given entirely by chiral loops, receives a large correction
at next-to-leading order.  Finally, we have also given
matching relations for our three-flavor representation to the $SU(2)$ case.

\newpage
%%%%%%%%%%%%%%%%%%%%%%%%%%%%%%%%%%%%%%%%%%%%%%%%%%%%%%%%%%%%%%%%%%%%%%%%%%%%%%%

\ack
We would like to thank Jambul Gegelia for useful discussions.
We acknowledge the support of the European Community Research Infrastructure
Integrating Activity ``Study of Strongly Interacting Matter''
(acronym HadronPhysics2, Grant Agreement n.~227431)
under the Seventh Framework Programme of the EU.
Work supported in part by DFG (SFB/TR 16, ``Subnuclear Structure of Matter''),
by the Helmholtz Association through funds provided to the virtual 
institute ``Spin and strong QCD'' (VH-VI-231), 
by BMBF ``Strong interaction studies for FAIR'' (grant 06BN9006),
and by the Bonn-Cologne Graduate School of Physics and Astronomy.

%\bibliographystyle{elsart-num}
%\bibliography{yyeft}
%\end{document}

\end{document}